\documentclass[review, times, sort&compress]{elsarticle}

\usepackage{hyperref}

\usepackage{amsmath}
\usepackage{amssymb}

\journal{Nuclear Instruments and Methods A}

\hypersetup{colorlinks=true, breaklinks=true, linkcolor=blue, menucolor=blue, urlcolor=blue}

\begin{document}

\begin{frontmatter}

\title{Response Functions for Detectors in Cosmic Ray Neutron Sensing}

\author[us,bo]{M. K\"ohli\corref{cor1}}
\ead{koehli@physi.uni-heidelberg.de}
\author[ms]{M. Schr\"on}
\author[us]{U. Schmidt}


\address[us]{Physikalisches Institut, Heidelberg University, Im Neuenheimer Feld 226, 69120 Heidelberg, Germany}
\address[bo]{Physikalisches Institut, University of Bonn, Nussallee 12, 53115, Germany}
\address[ms]{Helmholtz Centre for Environmental Research GmbH -- UFZ, Permoserstr 15, 04318 Leipzig, Germany}

\cortext[cor1]{Corresponding author}


\begin{abstract}
Cosmic-Ray Neutron Sensing (CRNS) is a novel technique for determining environmental water content by measuring albedo neutrons in the epithermal to fast energy range with moderated neutron detectors. We have investigated the response function of stationary and mobile neutron detectors typically used for environmental research in order to improve the model accuracy for neutron transport studies. Monte Carlo simulations have been performed in order to analyze the detection probability in terms of energy-dependent response and angular sensitivity for different variants of CRNS detectors and converter gases. Our results reveal the sensor's response to neutron energies from 0.1\,eV to $10^6$\,eV and highest sensitivity to vertical fluxes. The detector efficiency shows good agreement with reference data from the structurally similar Bonner Spheres. The relative probability of neutrons contributing to the overall integrated signal is especially important in regions with non-uniform albedo fluxes, such as complex terrain or heterogeneous distribution of hydrogen pools.

\end{abstract}


\end{frontmatter}


\section{Introduction}

Cosmic radiation is omnipresent on Earth and produces neutrons that interact with the atmosphere~\cite{ReitzCosmic} and the ground~\cite{Gordon2004}. In the last decades, many types of neutron detectors have been employed to observe those fluxes, such as high-energy neutron monitors~\cite{neutronmonitorAktuell} or Bonner Spheres~\cite{Bonner}. Neutrons in the epithermal to fast energy range (1\,eV to 10$^5$\,eV) are highly sensitive to hydrogen, which turns neutron detectors to efficient proxies for changes of environmental water content~\cite{Desilets2010}. The method of Cosmic-Ray Neutron Sensing (CRNS)~\cite{CRNS2008} is a promising tool for hydrological and environmental applications. The detector is usually mounted 1--2 meters above the ground surface, providing a significant exposure to far-field radiation from albedo neutrons which have further penetrated tens of decimeters into the soil~\cite{myself}. Hence, the neutron count rate is representative for the average root-zone water content in a footprint of several tens of hectares. The developments in the last years led to an enormous success of the method~\cite{Andreasen2017status} due to its large footprint, low maintenance, and non-invasive nature~\cite{Zreda2012}. To date, some features of the neutron response are still unknown and thereby introduce uncertainties to the measurement. For example, the influence of vegetation~\cite{Baatz2015}, detector location~\cite{notMyself2}, and cosmic-ray fluctuations~\cite{notMyself}. In the last few years, neutron simulations have been conducted to answer some of the open questions by transport modeling~\citep[e.g., ][]{myself, Andreasen2016, Schroen2017rover}. In order to reduce the enormous computational effort, which inevitably goes along with the large scale differences of a $\sim1$\,m$^3$ detector in a $\sim1\,$km$^3$ environment, effective response models have to be applied rather than using the geometrical detector itself in the simulation. However, such neutron detection models are sensitive to the specific response function of the detector. This property determines which neutron of which energy and from which direction is how likely to trigger an event. The current demand for autonomous techniques that monitor the water cycle is steadily increasing, while field measurements are required to get more and more accurate. It is thus important to understand the inherit sensitivity of the neutron detector to systematic effects in the environment, such as various hydrogen pools and cosmic-ray fluctuations. An accurate description of the detector's response function is a relevant step towards the goal to assess sensor signals with the help of simulations.

\section{Materials and Methods}

\subsection{The Cosmic Ray Neutron Probes}

\begin{figure}[ht!]
\centering
\includegraphics[width=\linewidth]{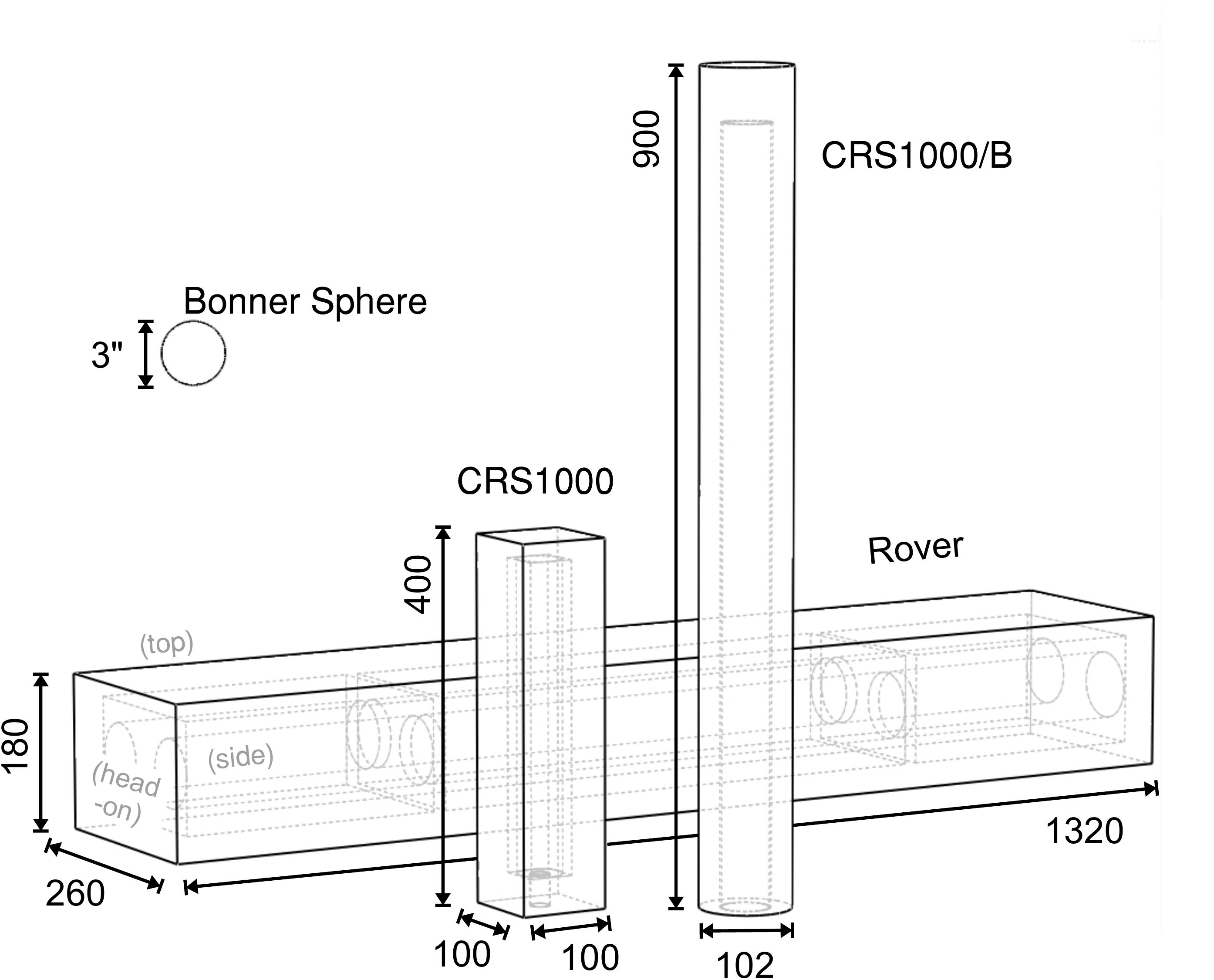}
\caption{Variants of the cosmic-ray neutron detectors modeled in this study. Dimensions are in units of millimiters. To scale a Bonner Sphere is illustrated in comparison.}
\label{fig:CRNSSensors}
\end{figure}

Cosmic-ray neutron sensors of type 'CRS'\footnote{Hydroinnova LLC, US} are commercially available in several configurations, spanning a variety of gaseous converters, geometry and orientation, see Fig.~\ref{fig:CRNSSensors}. The CRS1000 and CRS1000/B are mainly used in a stationary mode to monitor environmental neutron fluxes, and the Rover system is typically used in cars for mobile surveys of spatial neutron distributions. A description of the main components and detection physics can be found in~\cite{Zreda2012} and \cite{notMyself3}. The sensors comprise one or two moderated detector tubes sensitive to epithermal/fast neutrons, a high voltage generator, a pulse height analyzer, and a data logger with integrated telemetry. As a neutron moderator, high-density polyethylene of 1 inch thickness is used to encase the proportional counter tube that is filled with a neutron converter gas. The CRS1000 uses helium-3, while the CRS1000/B uses boron trifluoride, which requires larger detectors in order to achieve the same count rate due to its lower pressure and therefore absorption efficiency. The Rover is technically equivalent, but consists of significantly larger detectors than the stationary sensors and two tubes in one module to increase the detection rate, which in turn allows for higher temporal resolutions \citep[see e.g.][]{Chrisman2013, Schroen2017rover}. %

\subsection{URANOS}
The calculations presented here have been carried out using the recently developed code URANOS (\textbf{U}ltra \textbf{Ra}pid \textbf{N}eutron-\textbf{O}nly \textbf{S}imulation). The program, which is freely available online (\url{http://www.ufz.de/uranos}), has been designed as a Monte Carlo simulation of neutron interactions with matter. Recent applications cover studies of cosmic-ray induced albedo neutrons in environmental physics~\cite{myself, notMyself2, notMyself3, Schroen2017rover} and the characterization of neutron detectors for Spin Echo instruments in nuclear physics~\cite{myself2}. The standard calculation routine features a ray casting algorithm for single neutron propagation. The physics model follows the implementation declared by the ENDF database standard and was described by OpenMC~\cite{openmcRef}. It features the treatment of elastic collisions in the thermal and epithermal regime, as well as inelastic collisions, absorption and absorption-like processes such as evaporation. Cross sections, energy distributions and angular distributions were taken from the databases ENDF/B-VII.1~\cite{endfRef} and JENDL/HE-2007~\cite{jendlRef}.

\subsection{The Detector Model}

\begin{figure}[ht!]
\centering
\includegraphics[width=\linewidth]{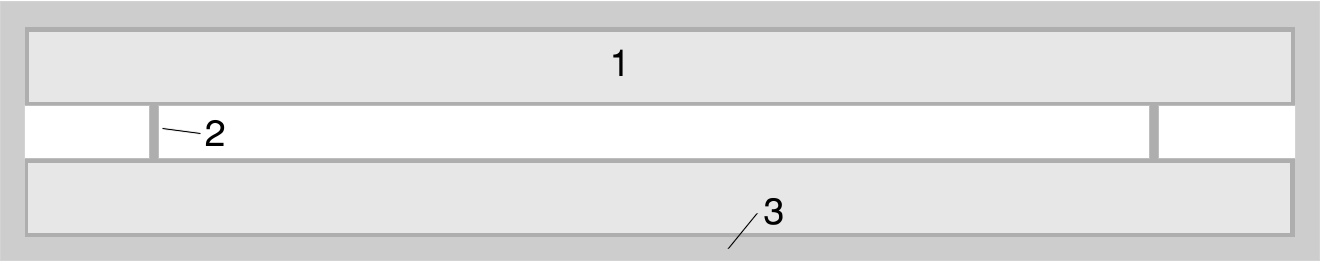}
\caption{Cross section of the  Rover detector simulation model with a length of 132\,cm and a width of 26\,cm. It features two gas filled proportional counter tubes in a stainless steel casing (1), aluminum mounting brackets (2) and a HDPE moderator (3).}
\label{fig:RoverModelTop2Name}
\end{figure}

URANOS handles model definitions by extruding voxels from layered images in a stack along the $z$ dimension. The central cutout of the detector configuration used in this study is shown exemplarily in Fig.~\ref{fig:RoverModelTop2Name}. The sensor geometry has been derived from actual devices and from supporting information provided by the manufacturer~\cite{hydroinnovaMan}, see also Fig.~\ref{fig:CRNSSensors}. Details of the mechanical parts have been reduced to features that have a significant influence on the neutron response, and only materials with significant macroscopic neutron cross sections have been considered. The materials used are: high-density polyethylene (CH$_2$) at a density of 0.98\,g/cm$^3$,
aluminum oxide (Al$_2$O$_3$) at 3.94\,g/cm$^3$, steel (Fe with 20\,\% Cr, 20\,\% Ni) at 8.03\,g/cm$^3$, boron trifluoride ($^{10}$B enriched BF$_3$ gas) at 2.76\,kg/m$^3$, $^{3}$He enriched noble gas at 0.125\,kg/m$^3$, and air (78\,\% N$_2$, 21\,\% O$_2$, 1\,\% Ar) at 1.2\,kg/m$^3$. The partial gas pressure has been set to 1.5\,bar for helium and to 0.5\,bar for boron trifluoride.

\begin{figure}[ht!]
\centering
\includegraphics[width=\linewidth]{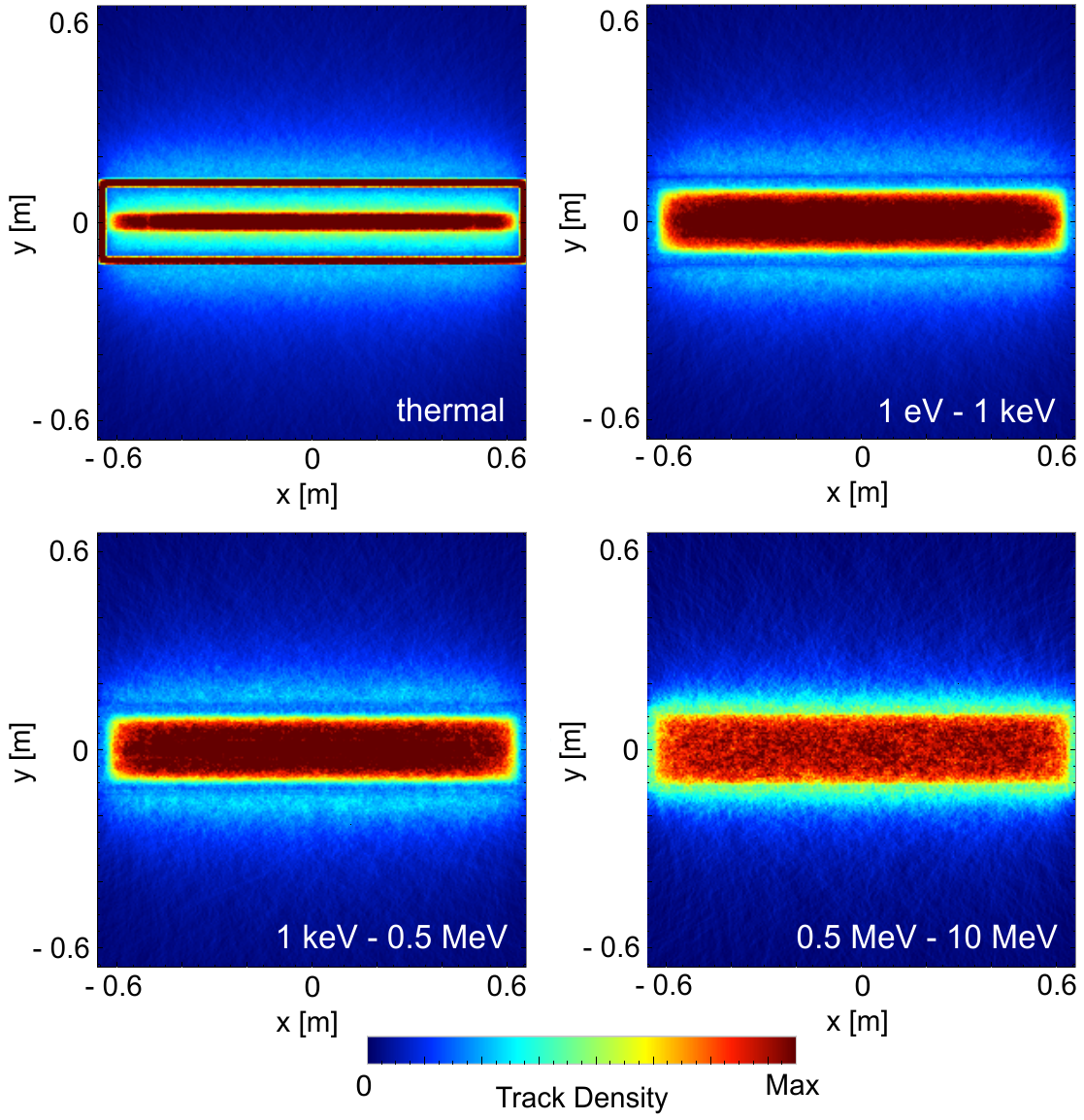}
\caption{Track density within the $^{10}$BF$_3$ Rover detector model using a randomly distributed flux from a plane source, illustrated for four energy regimes from thermal to MeV. The outer casing of the detector (see also Fig.\ref{fig:CRNSSensors}) consists of polyethylene, which becomes visible by the outward directed flux. Neutrons of high energies (lower right panel) do not undergo enough interactions to stay contained in the casing. Thermalized neutrons (upper left panel) are scattered within the moderator and are efficiently absorbed by one of the two tubes, with a probability of $\approx 0.5$ to be captured in either of them.}
\label{fig:RoverTop2}
\end{figure}

The orientation of the simulated detector tubes reflects the operational standard in environmental applications~\cite{CRNS2008}. The stationary systems (CRS1000 and CRS1000/B) are oriented upright, while the mobile system (Rover) is oriented horizontally. We further define two directions of the natural cosmic-ray neutron flux facing the 'top' and the 'sides' of the detector. Consequently, the 'top' facing flux runs from the surface upwards through the short cuboid face of the stationary sensor, and through the long cuboid face of the mobile detector. The 'side' facing fluxes run parallel to the surface through the long faces of the stationary detector and through two short and two long faces of the mobile detector.
In order to simulate incoming cosmic-ray flux from the atmosphere, neutrons were released randomly from a virtual plane of the same extension as the model dimensions. The number of neutrons absorbed in the converter gas divided by the total number of neutrons released is defined as the \emph{efficiency} of the setup, which intrinsically normalizes the efficiency to the detector area. The CRS1000/B consists of a proportional counter housed in a cylindrical casing. As here also the identical plane source definitions are used, this geometry leads to an ambiguity in the non form-fit efficiency definition, which has to be considered for interpreting the results, see also~\cite{McGregor2011167}. Nota bene: this definition is not phase space conserving under angular variation -- considering a neutron beam incident onto the sensor from a specific direction, the effective projected area of the corresponding cuboid face has to be taken into account. To study the behaviour of neutrons inside a detector system, the simulated neutron track density can provide insights. It is shown exemplarily for a $\mathrm{^{10}BF_3}$ Rover detector in Fig.~\ref{fig:RoverTop2}. The tracks represent $4 \cdot 10^7$ histories of incident neutrons with kinetic energies spanning 10 orders of magnitude. From the perspective of MeV neutrons, the path length through the polyethylene casing is in the order of the scattering length. This leads to an almost geometrically homogeneous distribution, where reflected neutrons have a negligible probability of reentry. Fast neutrons, $E \geq 1$\,keV, exhibit shorter scattering lengths while the energy lost by moderation allows for more neutrons in the boundary region to escape the device (seen by the 'glow' at the perimeter). For smaller energies, $E \leq 1$\,keV, the leakage out of the device is minimized while the number of interactions within the moderator is maximized. As soon as neutrons are thermalized, their absorption in the converter gas gets most effective.

\subsection{The energy response function}

Neutron detectors typically convert an incoming neutron flux to neutron count rates. However, not every particle of the incident neutron energy spectrum is counted as a detection event. The sensitivity of the detector can be different for different energies, depending on the conversion gas, the moderator type, and the geometrical configuration. The \emph{energy response} of a detector system, $R(E,\vartheta)$, quantifies this sensitivity as a function of neutron energy $E$ and incident angle $\vartheta$. It essentially maps the incoming neutron energy spectrum to a probability of counting an event. Previous studies, based on the modeling of Bonner Spheres~\cite{SphereExt}, showed that the detector response function can be approximated by the product of an energy-dependent efficiency term and an angular term:
\begin{equation}
R(E,\vartheta) = \epsilon (E) \cdot g(\vartheta)\,,
\label{eq:detResp}
\end{equation}
where an angle of $\vartheta=0 $ would correspond to an orthogonal neutron incidence. Averaged over the whole surface of the detector, the incoming flux can independently characterized as a function of these quantities.

\section{Results}

\subsection{Energy dependance}

\begin{figure}[ht!]
\centering
\includegraphics[width=\linewidth]{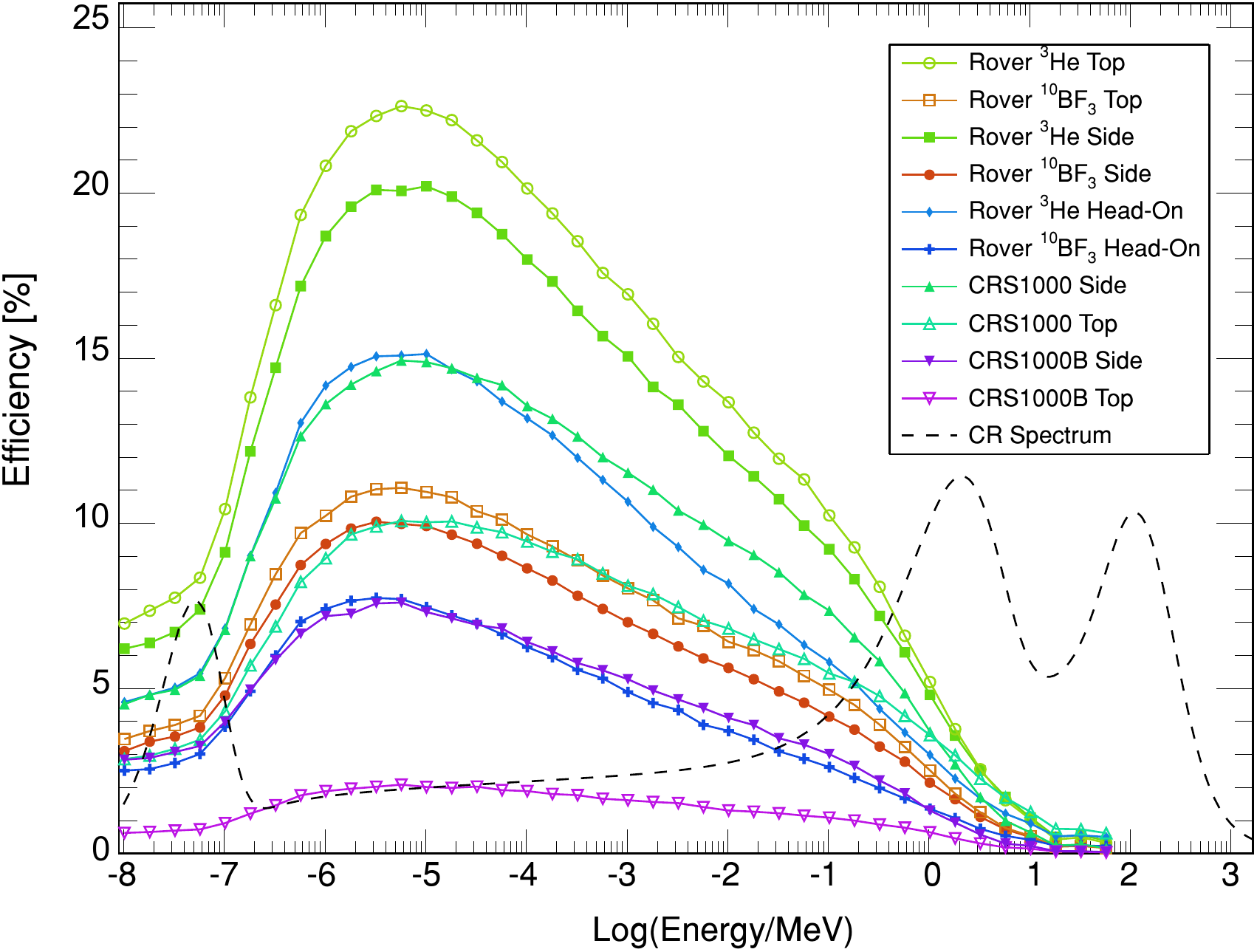}
\caption{Absolute counting efficiency for various actual cosmic-ray neutron sensing devices. The results for perpendicular irradiation are averaged over the entire surface for each setup. The cosmic-ray neutron spectrum from~\cite{Sato2006} illustrates the relative abundance of neutrons above the surface.}
\label{fig:CRNSEfficiencies}
\end{figure}

\begin{figure}[ht!]
\centering
\includegraphics[width=\linewidth]{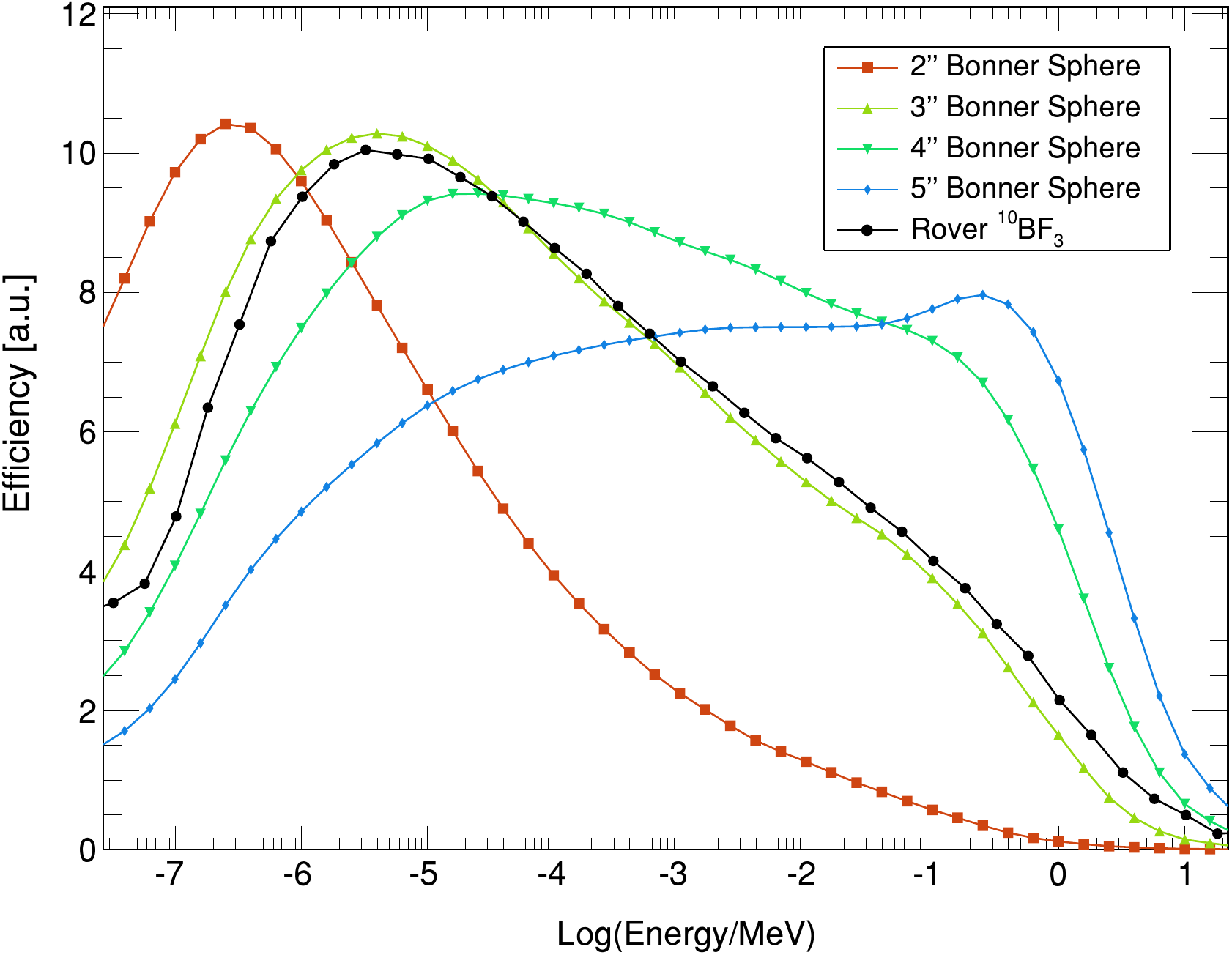}
\caption{Comparison of the energy dependent response function of different Bonner Spheres equipped with a 3.2\,cm spherical counter~\cite{Mares3He} and a cosmic-ray neutron detector. The efficiencies are scaled for quantitative matching. }
\label{fig:efficienciesBonnerComparison}
\end{figure}

The energy-dependent component of the neutron response, $\epsilon(E)$, has been calculated by URANOS simulations of different detector configurations. The results presented in Fig.~\ref{fig:CRNSEfficiencies} show that all detector models exhibit qualitatively similar energy response in the range from 0.1\,eV to 1\,MeV with a maximum between 1\,eV and 10\,eV. The main distinctions can be attributed to the absolute detection efficiency, which is a function of the detector model, the converter gas, and casing area. The latter is influenced by the geometry and orientation of the detector, as the surface neutron flux is averaged over the exposed area. Efficiencies scaled to the detector face area are provided in Table~\ref{tab:results}. Minor qualitative deviations of the response functions are noticeable for different aspect ratios in area coverage between moderator and counter tube, indicating that detector orientation is of minor importance for the energy efficiency, compare CRS1000 (top) and Rover (side) in Fig.~\ref{fig:CRNSEfficiencies}. The highest efficiency is achieved for neutrons in the energy range between 1\,eV and 100\,eV, while an average efficiency can be found between 0.1\,eV and 0.1\,MeV. The latter range corresponds to the 'water-sensitive domain' for the CRNS technique~\cite{myself}. The manufacturer has stated that the working energy range for the detectors is within 100\,eV to 10\,keV (unpublished data). 
This energy window appears to be too narrow compared to the results presented here, indicating a hitherto underestimation of near-thermal neutrons. A significant contribution of eV-neutrons was also suggested by other authors using empirical~\cite{McJannet2014} and modeling analysis~\cite{Andreasen2016}. The energy efficiency shows also remarkable similarity to reference curves of Bonner Spheres with equal moderator thickness, see Fig.~\ref{fig:efficienciesBonnerComparison}. As an example, the Rover detector system with the standard 1'' moderator thickness approximately corresponds to a 3'' moderator type with a 3.2\,cm spherical counter~\cite{Mares3He}, or to detectors equipped with a 4\,mm $^6$LiI crystal and a 2'' moderator~\cite{Mares2}. This example illustrates that the main influence on the energy-dependent response can be attributed to the thickness of the moderator. Similar results also have been presented for portal monitor type detectors~\cite{GEANTComparisonDetector}. For the actual integration of such a response into an environmental neutron transport model a function derived from a cubic spline interpolation of this data or Bonner Sphere calculations can be utilized.

\subsection{Angular dependance}

\begin{figure}[ht!]
\centering
\includegraphics[width=\linewidth]{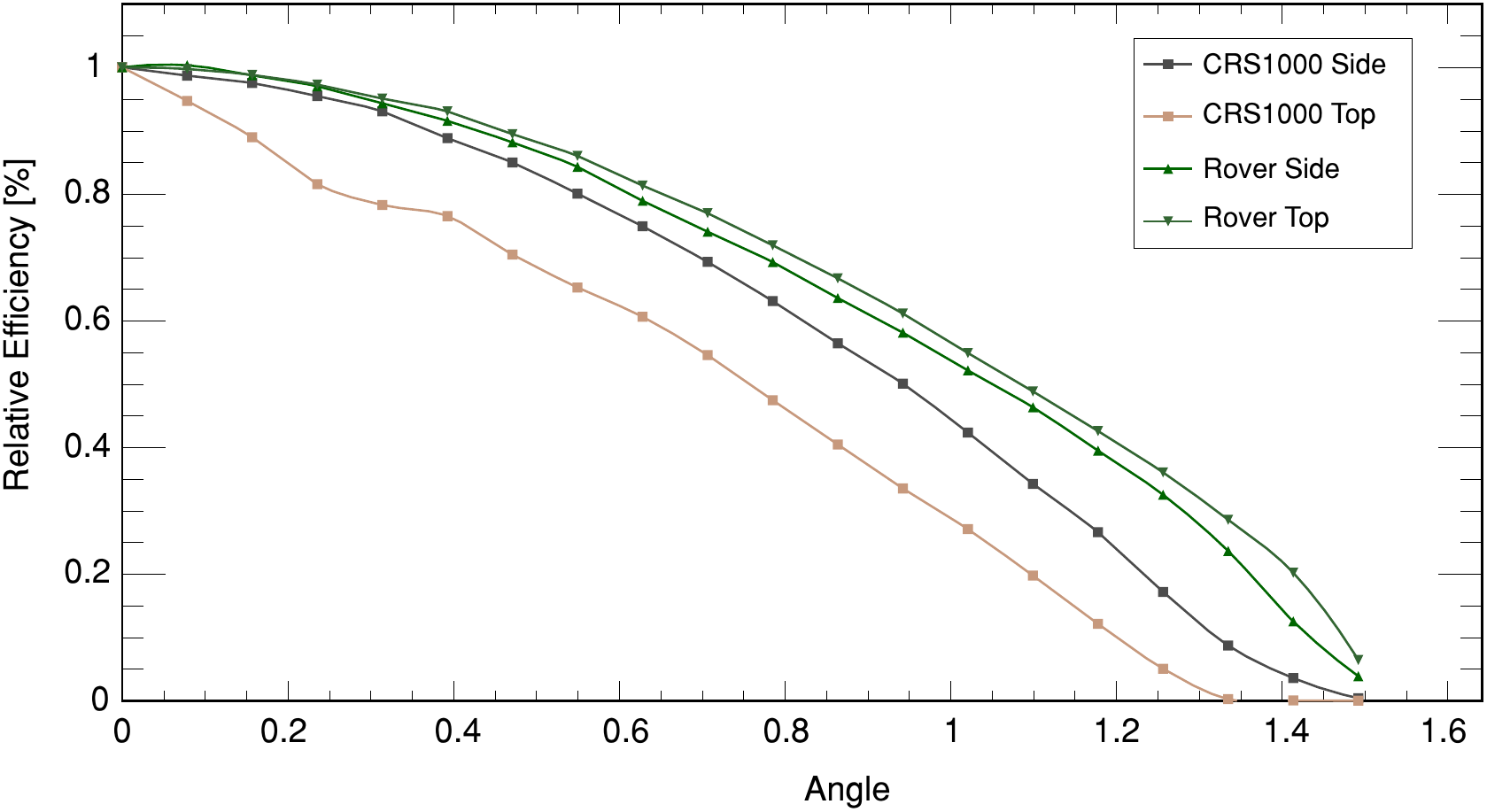}
\caption{Relative counting efficiency for some selected detector models and faces as a function of inclination angle of the incoming flux representing the angular term of~\eqref{eq:detResp}. }
\label{fig:RelEfficiency}
\end{figure}

The angular sensitivity component of the response function, see~\eqref{eq:detResp}, is shown in Fig.~\ref{fig:RelEfficiency}, averaged over the all energies and the detector face area. The angle $\vartheta=0$ is oriented perpendicular to the surface and efficiencies are normalized to $g(\vartheta=0)=1$. The detectors show lowest sensitivity to neutrons from directions parallel to the surface, as for slant angles ($\vartheta=\pi/2$ or $90^\circ$) the probability of detection drops to zero. Highest sensitivity for all detectors is achieved for orthogonal incidence with $\vartheta=0$. This result stresses the importance of accounting for neutron fluxes directly from atmospheric cosmic rays~\cite{notMyself} and directly from beneath the sensor~\citep{Schroen2017rover}. An analytical approximation of the angular distribution can be given as:
\begin{equation}
g(\vartheta) = 1.24 - 0.254\,e^{1.087\,\vartheta}.
\label{eq:angulareffDet}
\end{equation}
Like for the energy-dependent term discussed above, the geometric arrangement of the moderator has a minor influence on the response function.

\subsection{Detection probability within the case}

\begin{figure}[ht!]
\centering
\includegraphics[width=\linewidth]{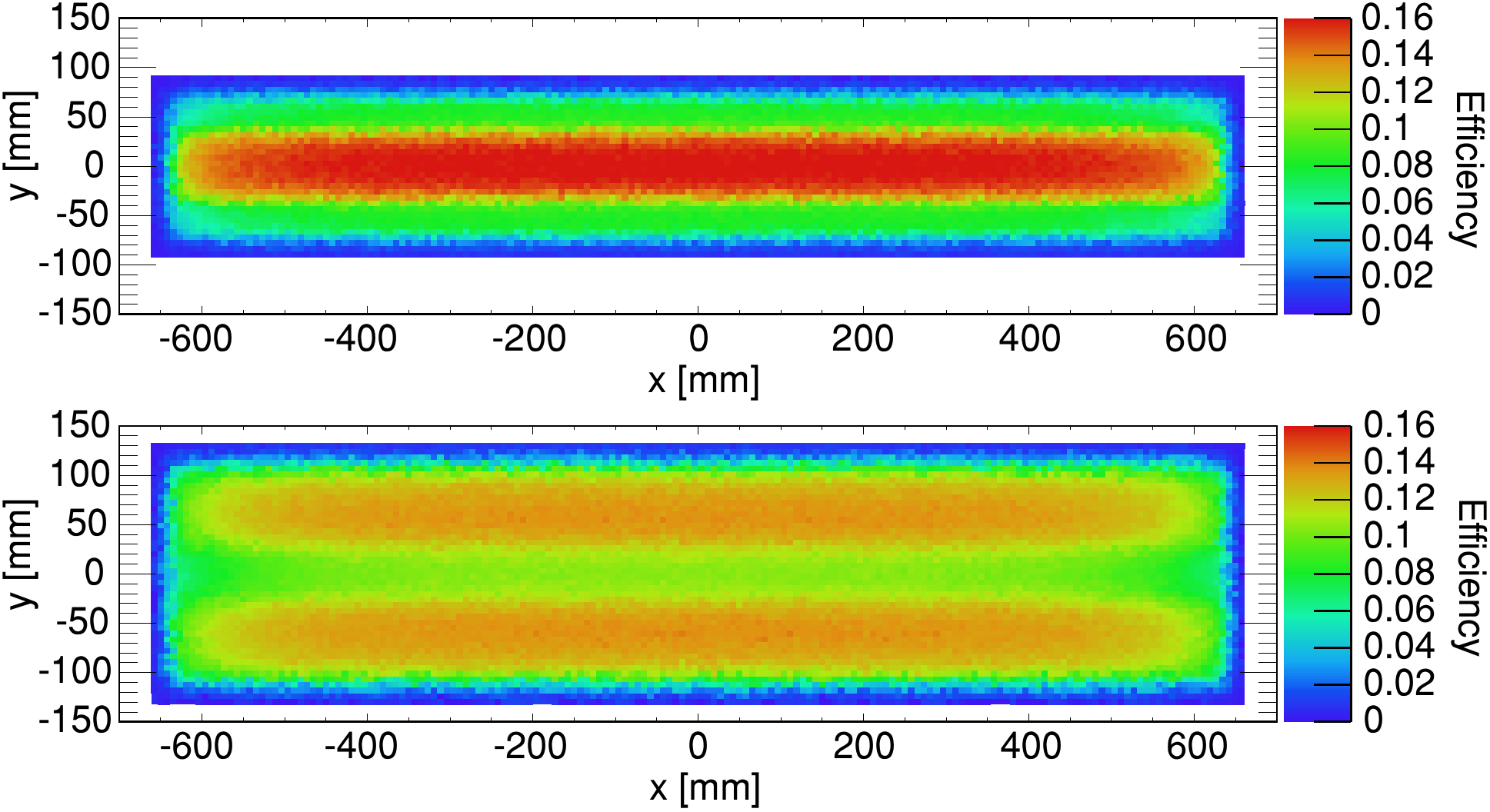}
\caption{Efficiency map for the $^{10}$BF$_3$ Rover system for orthogonally incident neutrons from the side (top panel) and the top direction (bottom panel), depicting the probability of being absorbed in the converter dependent on the $x,y$ coordinate entering the detector.}
\label{fig:RoverSpatialEfficiency}
\end{figure}

The results above have addressed the detector efficiency averaged over the entire detector surface. However, the detector case itself cannot be considered as a homogeneously responsive device. A neutron hitting the detector centrally has a much higher absorption probability than a neutron entering at the very edge of a moderator. The spatial distribution of the detector efficiency can be illustrated with an efficiency map. As an example, Fig.~\ref{fig:RoverSpatialEfficiency} shows the boron trifluoride Rover system (two tubes) for 10\,eV neutrons from the side- and top-facing perspective. The color scale represents the detection probability for a normally incident neutron, showing that detection is more probable but in a narrower area for sideways incident neutrons compared to topwards incident neutrons. Although Fig.~\ref{fig:RoverModelTop2Name} showed for the epithermal/fast regime a homogeneous distribution of the tracks inside the casing, provided the individual original impact location the efficiency varies significantly. 

\subsection{Uncertainty Analysis}

Simulations performed in this study were conducted with $10^6$ released neutrons, which corresponds to a relative statistical error of the detector response $R$ of $s_R = 10^{-2}/\sqrt{R}$, where $R=R(E,\theta)$ is given in units of percent and usually stays below 1\,\%. The good agreement with reference calculations from literature confirmed the reasonability of our approach (Fig.~\ref{fig:efficienciesBonnerComparison}). Systematic errors of potential relevance mainly involve the assumptions on material composition and geometry. For polyethylene the scattering kernel was emulated by water, which, due to the higher mobility of water molecules, could have biased the resulting efficiency by up to 10\,\% particularly in the thermal regime. The fabrication related variations of polyethylene density could further alter the macroscopic cross sections of the real detector in the order of (1--2)\,\%, thereby shifting the actual response function towards thicker or thinner moderators.  Moreover, the abstraction level used for the modeled detector geometry has been high, as only moderator, absorber, and the metal parts have been taken into account. Nonetheless, our calculations showed that even drastic changes of the arrangement had only marginal effects on the response function.

\section{Discussion}

\subsection{Pseudo efficiency}

For each detector model a pseudo efficiency value $\epsilon^*$ can be calculated, which denotes the maximum efficiency normalized by the facing area of the device. For the total surface area, this quantity can be interpreted as a measure for the total count rate in an homogenous fast neutron field. Tab.~\ref{tab:results} summarizes the response values for all detector models studied in this work. The small, vertical stationary probes show similar efficiency for both sensor types, as the larger size of the boron trifluoride tube compensates the lower total cross compared to helium-3. The Rover detector types have been assumed to have the same size, leading to helium-3 gas efficiencies higher by a factor of 2 compared to boron-10.

%
\begin{table}[ht]
\begin{center}
\begin{tabular}{lccrcl}
model	& gas & face		& $\epsilon_\text{max}$ & area [m$^2$] & $\epsilon^*$\\
\hline
CRS1000 &$^3$He & top     & 10.0\,\% & 0.010 & 0.100  \\
CRS1000 &$^3$He & side    & 14.9\,\% & 0.040 & 0.596  \\
CRS1000/B &$^{10}$BF$_3$& top     & 2.1\,\%  & 0.009 & 0.018  \\
CRS1000/B &$^{10}$BF$_3$& side    & 7.6\,\%  & 0.090 & 0.684  \\
Rover &$^{10}$BF$_3$ & top     & 11.1\,\% & 0.343 & 3.810  \\
Rover &$^{10}$BF$_3$ & side    & 10.0\,\% & 0.238 & 2.376  \\
Rover &$^{10}$BF$_3$ & head & 7.7\,\%  & 0.047 & 0.360  \\
Rover &$^3$He        & top     & 22.6\,\% & 0.343 & 7.760  \\
Rover &$^3$He        & side    & 20.2\,\% & 0.238 & 4.800  \\
Rover &$^3$He        & head & 15.1\,\% & 0.047 & 0.710  \\
\hline
CRS1000   &$^3$He & all     & --       & 0.180 & 2.584  \\
CRS1000/B &$^{10}$BF$_3$& all     & --       & 0.377 & 2.772  \\
Rover &$^{10}$BF$_3$ & all     & --       & 1.256 & 13.09  \\
Rover &$^3$He        & all     & --       & 1.256 & 26.54  \\
\hline
\end{tabular}
\end{center}
\caption{Summary of the detector models, the facing direction of the examined neutron flux, their maximum efficiency, the area exposed to the neutron flux, and the pseudo efficiency, $\epsilon^*=\epsilon_\text{max}/\text{area}$, denoting a measure for the total count rate.}
\label{tab:results}
\end{table}

\subsection{Implications of the angular sensitivity}

Separate analysis of background and albedo neutrons has not been performed in this work, as the spatial and angular distributions strongly depend on the reflectivity of the ground, which is a function of snow or soil water, and on the sensor height above the interface. Nonetheless, our results shed light on effects which were actually observed in the field. Several researchers reported preliminary results that indicate higher sensitivity to local effects for horizontal detectors compared to vertical ones, which is observable for example by the change of road materials or nearby water bodies in roving applications. In theory, the neutron flux exhibits a vertical symmetry due to a strong suppression of the horizontal phase space by the ground interface. Given the effective detection area of the cuboid sensor and its angular response, both conformations do exhibit a different sensitivity to the neutron flux directly below the sensor. In other words, near-field effects, like reported in~\cite{notMyself2} and \cite{Schroen2017rover}, can be attributed to the relatively high sensitivity to the downward direction and the flux distribution around the detector. Although the internal assembly of the detector is visible in the impact location map, Fig.~\ref{fig:RoverSpatialEfficiency}), the device is small compared to the diffusion length of environmental neutron fluxes \cite{DesiletsZreda2013, myself}. Therefore, higher accuracy of the presented computational results would not lead to relevant information for environmental research.

\subsection{Implications of the energy sensitivity}

We found a significant mismatch between the operational energy range given usually referenced in the literature of (0.1--10)\,keV and the results shown in this work. Yet, the implications are probably moderate. The entire spectrum from 1\,eV to 0.1\,MeV is dominated by elastic scattering and the cosmic-ray induced density of albedo neutrons. It is related to the environmental water content and scales uniformly in this regime \cite{myself}. Hence, the asymmetric shape of $\epsilon(E)$ has a minor influence on the sensitivity of the device in terms of soil moisture sensing. According to additional simulations by the author (not shown), the change of the sensor's footprint radius is negligible if the revised response function was accounted for (compare also \cite{myself}). The similarity to the response of Bonner Spheres, however, explains the reported influence of thermal neutrons to the CRNS detectors~\cite{McJannet2014, Andreasen2016}. A slightly thicker moderator would improve the performance of the devices by shifting the response function out of the thermal regime, while an absorber shield around the casing would result in a nearly complete suppression of the thermal neutron leakage into the device.
The overall neutrons detected by the sensor can be calculated by the product of the response function and neutron energy spectrum at the surface, compare also Fig.~\ref{fig:CRNSEfficiencies}. This energy-dependent efficiency can be used to estimate the relative contribution of neutrons of a specific type to the sensor signal. From this it can been derived that the contribution from 'thermal' neutrons below the cadmium cutoff at 0.5\,eV is about (15--22)\,\% and from neutrons above 0.1\,MeV about 
(18--25)\,\%, considering that such numbers highly depend on the detector configuration, moderator thickness, and the soil chemistry.

\section{Summary and Outlook}

In this paper we analyzed the response functions of cosmic-ray neutron sensors in terms of energy-dependent detection efficiency and angular sensitivity. The investigated detectors comprise vertical (CRS1000) and horizontal tubes (Rover), which are specifically manufactured for environmental and hydrological research, each moderated by 1 inch of polyethylene and filled with either $\mathrm{^{10}BF_3}$ or $\mathrm{^3He}$. Simulations of the neutron response have been conducted using the neutron transport code URANOS. The results show that the energy window of highest response ranges from 0.1\,eV to $10^6$\,eV and peaks between 1\,eV and 10\,eV. Hence, a significant fraction of neutrons are contributing to the sensor signal below and above the hitherto accepted range of ($10^2$ to $10^4$)\,eV. The simulations agree well with the response functions from Bonner Spheres of equally large diameter. The angular distribution of incoming fluxes indicates a prominent sensitivity to incident neutrons from the vertical direction. Hence, the consideration of background cosmic-ray fluxes as well as albedo neutrons from the ground below the sensor should be of particular importance for environmental applications of these neutron detectors.

\section*{Acknowledgments}
URANOS was developed by the project 'Neutron Detectors for the MIEZE method', funded by the German Federal Ministry for Research and Education (BMBF), grant identifier: 05K10VHA. Part of the funding was provided by the German Federal Ministry for Research and Education (BMBF) in the frame of the project 'Forschung und Entwicklung hochaufl\"osender Neutronendetektoren', grant identifier: 05K16PD1.
The research was supported by the Terrestrial Environmental Observatories (TERENO), which is a joint collaboration program involving several Helmholtz Research Centers in Germany.

\section*{References}
\biboptions{ sort&compress}


\end{document}